\documentclass[useAMS,usenatbib]{mn2e}
\usepackage{latexsym,graphicx,natbib,amssymb}

\title[The influence of gas expulsion on the evolution of star clusters]
 {A comprehensive set of simulations studying the influence of gas expulsion on star cluster evolution}
      
\author[H. Baumgardt, P. Kroupa]
{
  H. Baumgardt\thanks{e-mail: holger@astro.uni-bonn.de (HB);
    pavel@astro.uni-bonn.de (PK)}, 
  and P. Kroupa\\ 
  Argelander-Institut f\"ur Astronomie, Universit\"at Bonn, Auf dem H\"ugel 71, 53121 Bonn,
  Germany\\
}

\begin{document}

\date{Accepted ????. Received ?????; in original form ?????}

\pagerange{\pageref{firstpage}--\pageref{lastpage}} \pubyear{2006}

\maketitle

\label{firstpage}

\begin{abstract}
We have carried out a large set of $N$-body simulations studying the effect of
residual-gas expulsion on the survival rate and final properties of star clusters.
We have varied the star formation efficiency, gas expulsion timescale and strength of 
the external tidal field, obtaining a three-dimensional grid of models which can be used to
predict the evolution of individual star clusters or whole star cluster systems by
interpolating between our runs. The complete data of these simulations is made 
available on the Internet.
 
Our simulations show that cluster sizes, bound mass fraction and velocity profile are
strongly influenced by the details of the gas expulsion. Although star clusters can 
survive
star formation efficiencies as low as 10\% if the tidal field is weak and the gas is 
removed only slowly, our simulations indicate that most star clusters are destroyed or
suffer dramatic loss of stars during the gas removal phase. Surviving clusters have 
typically expanded by a factor 3 or 4 due to gas removal, implying that star clusters 
formed more concentrated than as we see them today. Maximum expansion factors seen
in our runs are around 10. If gas is removed on timescales smaller than the initial crossing time, 
star clusters acquire strongly radially anisotropic velocity dispersions outside their
half-mass radii. Observed velocity profiles of star clusters can
therefore be used as a constraint on the physics of cluster formation. 
\end{abstract}


\section{Introduction}
\label{sec:intro}

It is well known that most, if not all, stars form in star clusters. 
Star clusters can therefore be viewed as the fundamental building blocks of galaxies since the
galactic disc velocity distribution \citep{k02, k05}, the galactic stellar
mass function \citep{wk05} and the galactic-field binary-star population \citep{k95} are all established
by them. 

Star clusters form as so
called embedded clusters within the dense cores of giant
molecular clouds. The star formation efficiency (SFE), i.e. the fraction of gas
that is converted into stars, can be defined as follows:
\begin{equation}
\epsilon = \frac{M_{ecl}}{M_{ecl}+M_{gas}}
\end{equation}
where $M_{ecl}$ is the total mass of stars formed in the embedded cluster and $M_{gas}$ the mass of the gas not
converted into stars. Inside molecular cloud cores, the star formation 
efficiency is usually smaller than $\epsilon<30$\% 
\citep{l99, ll03}. Over the whole molecular cloud, the star formation efficiency is
even lower and only of the order of a few percent \citep{cb04}.
Since the time-scale for the formation of a star cluster is $\sim 10^6$ yrs, far larger than the
typical crossing time-scale of a star cluster, star clusters are most likely in virial 
equilibrium before the gas is lost, and the velocity of the stars reflects the overall 
gravitational potential \citep{k05}. Together with the low SFEs, this implies that once the primordial 
gas is expelled by UV radiation and massive stellar winds from OB stars or supernova
explosions, star clusters will become super-virial and their further dynamical evolution
will be strongly affected by the gas loss.

The impact of gas expulsion and the subsequent response of the star cluster has
been the subject of a large number of theoretical investigations 
\citep{tut78, hill80, lmd84, good97a, good97b, kpm99, a00, kah01, gb01, bk03a, bk03b, fk05, bg06}. 
Using the 
virial theorem, \citet{hill80} argued that if the gas is lost instantaneously and 
the SFE is below 50\%, the 
entire cluster should be disrupted. Later analytic and $N$-body modeling 
\citep{lmd84, gb01, bk03a, bk03b} have mainly confirmed Hills ideas but also shown that the 
actual limit for bound star cluster 
formation in case of instantaneous gas-loss is somewhat lower. \citet{lmd84} obtained a bound
star cluster with a SFE as low as $\epsilon=40$\%, and the simulations of \citet{gb01}
indicate that the critical SFE is below $35$\%. Using more realistic assumptions, \citet{kah01}
computed the evolution of Orion-Nebula-like clusters showing that they disperse leaving young
Pleiades pre-cursors embedded in expanding OB associations.

In the early phases of the evolution, before the first supernova explosions go off,
gas is lost from star clusters mainly through radiation pressure and stellar winds
from bright OB stars. Both radiation pressure and stellar
winds need some time to expel the gas from a star cluster. A single 10 M$_\odot$ (85 M$_\odot$)
star for example ejects an energy of $3 \cdot 10^{50}$ ($3 \cdot 10^{51}$) erg/s into the 
interstellar medium in the form of radiation and mechanical energy \citep{k05}. This is enough to unbind 
gas from a $10^4 M_\odot$, $r_h=1$ pc cluster within
$\approx 10^5$ yrs, comparable to the crossing time of such a cluster. 

If the gas is not lost instantaneously but adiabatically, i.e. over a timescale much longer
than the cluster's crossing time, the cluster stars
can adjust to the change in potential and the critical SFE needed for star cluster 
survival can be significantly lower. \citet{lmd84} obtained bound clusters with
SFEs as low as 20\% if the gas expulsion timescale was equal to several initial crossing
times. They also found that between 10\% to 80\% of the initial stars were lost
from such systems, especially stars from the outer cluster parts where the initial
velocities are more likely to exceed the escape velocities of the cluster once
the gas is expelled. Similarly, \citet{good97a} found that star clusters in 
tidal fields can form bound systems with SFEs as low as 20\% which will survive for 
a Hubble time if the initial concentration of the cluster is high enough.
Star clusters might also survive low SFEs if they form through merging of individual 
star clusters in star cluster complexes and the gas is lost in different clusters 
at different times \citep{fk05}. Finally, \citet{a00} studied cluster response to gas
blow out assuming that the SFE varies with radius, finding core survival for high
central values of the SFE.

While many aspects of the residual gas expulsion have already been studied in the
literature, nobody has so far tried to perform a systematic study of the survival
limit and the final parameters of star clusters evolving under residual gas expulsion.
In addition, with the exception of \citet{good97a} and \citet{kah01}, the influence of the external tidal 
field has so far been largely neglected. 

In the present paper, we have performed
a large parameter study of residual gas expulsion from star clusters, varying the
star formation efficiency, the ratio of the gas expulsion time scale to the crossing 
time of the star cluster and the ratio of the half-mass radius to the tidal radius
of the star cluster. Our grid of models will be useful in later studies of individual
star clusters and also whole star cluster systems since the evolution of the clusters 
can be determined by interpolation between our grid points, without the need for 
further simulations. This makes it possible to determine the effect of gas expulsion
on whole cluster systems where the large number of clusters prevents simulations
for all individual clusters. 

Our simulations rely on a number of simplifying assumptions, like e.g. that
all stars in a cluster form at the same time, that the SFE is constant with radius,
or that the overall gas density
is spherical. However, relaxing most of these assumptions should make relatively
little change to our results. For example, whether stars in a cluster form at the 
same time or not has no influence on the dynamical reaction of the cluster to gas expulsion
since this reaction depends only on the fraction of gas thrown out. What
happens to the gas which remains, whether it has been turned into stars
10 Myrs before gas expulsion, or 1 Myr before, or at the time the other gas
leaves is not relevant. Similarly, gas might leave a cluster in certain parts, where
e.g. the overall density is lower, earlier than in other parts, leading to a non-spherical
distribution of the gas. However the dynamical influence of the gas is not through its 
density distribution but through the potential this density gives rise to and a
potential field is always much smoother than the underlying density. 
Hence, the assumption that the external gas distribution is spherical should
also have only a small influence on our results.

Our paper is organised as follows: In Sec.\ \ref{sec:Nbody}
we describe the initial set-up of our models and in Sec.\ \ref{sec:results} we present
the main results. Sec.\ref{sec:concl} finally presents our conclusions.

\section{The models}
\label{sec:Nbody}

In our runs, we assumed that the SFE does not depend on the 
position
inside the cluster, so gas and stars followed the same density distribution initially,
which was given by a Plummer model. The gas was not simulated directly, instead its influence
on the stars was modeled as a modification to the equation of motion of stars. 
Modeling the gas as an additive potential has been shown by \citet{gb01} to be a good approximation 
of the essential physics driving early cluster evolution.
We used the collisional $N$-body code NBODY4 \citep{a99} to perform the simulations. Since
NBODY4 uses a Hermite scheme to integrate the motion of stars, the correction terms due to
the external gas on the acceleration and its first derivative have to be evaluated at each step.
For a Plummer model, the necessary correction
terms can be derived analytically and are given by:
\begin{eqnarray}
\nonumber \frac{d^2 \vec{r}}{d t^2} & = & - \frac{G \; M_{gas}(t)}{\left(r^2+r_{Pl}^2\right)^{1.5}} \; \vec{r} \\
\nonumber  \frac{d^3 \vec{r}}{d t^3} & = & - \frac{G\; \dot{M}_{gas}(t)}{\left(r^2+r_{Pl}^2\right)^{1.5}} \vec{r}
- \frac{G \; M_{gas}(t)}{\left(r^2+r_{Pl}^2\right)^{1.5}} \dot{\vec{r}} \\
 & &  + 3 \frac{G \; M_{gas}(t)}{\left(r^2+r_{Pl}^2\right)^{2.5}} \; (\vec{r} \dot{\vec{r}}) \; \vec{r} \;\; ,
\end{eqnarray}
where $\vec{r}$ is the position vector of a star relative to the cluster centre, $G$ the gravitational constant,
$r=|\vec{r}|$, $M_{gas}(t)$ is the 
total mass in gas left at time $t$ and $r_{Pl}$ is the scale length of the Plummer model. 
Gas expulsion was assumed to start at a certain time $t_D$, which
was set equal to one $N$-body time unit \citep{hm85}, equivalent to $1/\sqrt{8}$ of a crossing
time at the clusters virial (= gravitational) radius \citep{bt87}. This
small offset was introduced in order to test for each cluster if our set-up program created clusters 
that start in an equilibrium state when the external gas is present. 
After the delay
time $t_D$, the gas density was decreased exponentially on a characteristic time $\tau_M$, so the 
total gas left at later times is given by \citep{kah01}:
\begin{equation}
 M_{gas}(t) = M_{gas}(0) \; e^{-(t-t_D)/\tau_M} \; \; .
\end{equation}
$\tau_M$ will henceforth be called the gas expulsion timescale. 
The influence of the external tidal field was modeled in the so-called 'Near Field
Approximation', which assumes that the size of the star
cluster is much smaller than its distance from the galactic centre. For clusters moving
in a circular orbit through a spherical galactic potential, the equation of motion of stars in a reference frame rotating with 
the star cluster can be expressed as \citep{a85}:
\begin{equation}
\frac{d^2 \vec{r}}{d t^2} = \left. \frac{d^2 \vec{r}}{d t^2}\right|_{Cl} \!\! - 2 \vec{\omega} \times \frac{d \vec{r}}{d t} + \omega^2 (3 x \vec{e}_x - z \vec{e}_z) \;\; .
\end{equation}
Here the first term on the right hand side is the gravitational acceleration due to the star cluster
(both stars and gas), the second term is the Coriolis acceleration and the third term is a combination
of centrifugal and tidal forces and it was assumed that the cluster moves in the x-y plane. The 
angular velocity $\omega$ is given by:
\begin{equation}
\omega= \sqrt{\frac{G \; M_G}{R^3_G}}
\end{equation}

For real star clusters, six parameters will 
determine their fate under the influence of gas expulsion: the total initial cluster mass $M_{ecl}+M_{gas}$, the 
half-mass radius, $r_h$, and galactocentric distance, $R_G$, of the cluster, the total mass, $M_G$, of the parent galaxy inside the 
cluster position, and the star formation efficiency, $\epsilon$, and gas expulsion 
timescale, $\tau_M$. The number of parameters can however be reduced efficiently: The timescale of gas expulsion 
for example enters not through its absolute value, but only through its ratio with the initial 
crossing time $t_{Cross}$ of the star cluster. At the virial radius, $r_v$, the initial crossing time 
$t_{Cross}$ is given by \citep{hh03}: 
\begin{equation}
 t_{Cross} = 2.82 \frac{r^{1.5}_v}{\sqrt{G} \; \sqrt{M_{gas}+M_{ecl}}}
\end{equation}
If gas expulsion happens slowly, stars can adjust to the change in potential
and clusters expand adiabatically. In the other extreme, gas is lost instantaneously and clusters are 
strongly effected by gas expulsion. Similarly, $R_G$ and $M_G$ determine together with
the cluster mass the tidal radius of the star cluster: 
\begin{equation}
 r_t = \left( \frac{G \; M_{ecl}}{3 \; M_G} \right)^{1/3} R_G  \;\; .
\end{equation}
Dynamically important is not the absolute 
value of $r_t$, but only the ratio of
$r_h/r_t$: If this ratio is low, clusters are nearly isolated and can expand freely, while if the ratio
is high, they are strongly tidally limited and easily destroyed. 

Hence, within the framework of our 
model, the fate of a star cluster can
be deduced by specifying only three parameters: The star formation efficiency $\epsilon$, the ratio of the 
gas expulsion time scale to the crossing time of the star cluster, and the strength of the 
external tidal field, quantified by the ratio of the half-mass radius to the tidal radius.

This reduction in the number of parameters 
makes it feasible to run a grid
of models covering the complete parameter space. To this end, we have performed a set of $N$-body 
simulations, varying
the initial star formation efficiency, the ratio of half-mass radius to the tidal radius $r_h/r_t$,
and the ratio of the gas expulsion time scale to the crossing time of the cluster $\tau_M/t_{Cross}$.
The following values were chosen as grid-points:\\[+0.3cm]
\noindent
\begin{tabular}{l@{\hspace*{0.1cm}}l}
$\epsilon$: & 5\%, 10\%, 15\%, 20\%, 25\%, 33\%, 40\%, 50\%, 75\%  \\[+0.1cm]
$r_h/r_t$: & 0.01, 0.033, 0.06, 0.1, 0.15, 0.2 \\[+0.1cm]
$\tau_M/t_{Cross}$: & 0.0, 0.05, 0.10, 0.33, 1.00, 3.0, 10.0\\[+0.3cm]
\end{tabular}

The chosen range of parameter values essentially covers the relevant part of parameter
space since star formation efficiencies of less than 10\% lead to cluster destruction
while efficiencies higher than 75\% have not been seen in nature and lead to nearly complete
cluster survival. Also, clusters with $r_h/r_t=0.01$
are nearly isolated and we do not expect that results will change much when choosing even
smaller ratios, while clusters with $r_h/r_t=0.2$ are strongly limited by the tidal field and
nearly all of them are destroyed. Similarly, our values of $\tau_M/t_{Cross}$ cover the 
interesting range for open/globular clusters. 

All runs were performed with the collisional $N$-body code NBODY4
\citep{a99} on the GRAPE6 computers \citep{mfkn03} of our group at Bonn University.
All simulated clusters contained 20.000 equal-mass stars initially,
distributed according to a Plummer sphere. The simulations were
run for 1000 initial $N$-body times (equivalent to about 300 initial crossing times). We found that this was 
sufficient since the gas is lost on a much shorter timescale and all clusters have settled
into an equilibrium state by this time. Also, running for longer times would have meant
that relaxation effects could have become important since the relaxation
time of the stellar component is about 250 initial crossing times. Our models can be viewed as a first step 
towards realistic models since two-body relaxation is not important in our runs and stellar 
evolution is neglected completely. In realistic star clusters, both effects would become 
important in the later stages of cluster evolution.

\begin{table*}
\caption[]{Details of the $N$-body runs which lead to the formation of a bound cluster.}
\begin{tabular}{c@{\hspace*{0.3cm}}c@{\hspace*{0.3cm}}c@{\hspace*{0.3cm}}c@{\hspace*{0.3cm}}c@{\hspace*{0.3cm}}r@{\hspace*{0.6cm}}c@{\hspace*{0.3cm}}c@{\hspace*{0.3cm}}c@{\hspace*{0.3cm}}c@{\hspace*{0.3cm}}c@{\hspace*{0.3cm}}r@{\hspace*{0.6cm}}c@{\hspace*{0.3cm}}c@{\hspace*{0.3cm}}c@{\hspace*{0.3cm}}c@{\hspace*{0.3cm}}c@{\hspace*{0.3cm}}r}
\noalign{\smallskip}
 SFE &  $r_h/r_t$ & \multicolumn{1}{c}{$\frac{\tau_M}{t_{Cr}}$} & $\frac{M_{*f}}{M_{ecl}}$ & \multicolumn{1}{c}{$\frac{r_{hf}}{r_{hi}}$} & $\beta_v$$\;\;$ & 
 SFE &  $r_h/r_t$ & \multicolumn{1}{c}{$\frac{\tau_M}{t_{Cr}}$} & $\frac{M_{*f}}{M_{ecl}}$ & $\frac{r_{hf}}{r_{hi}}$ & $\beta_v$$\;\;$ & 
 SFE &  $r_h/r_t$ & \multicolumn{1}{c}{$\frac{\tau_M}{t_{Cr}}$} & $\frac{M_{*f}}{M_{ecl}}$ & $\frac{r_{hf}}{r_{hi}}$ & \multicolumn{1}{c}{$\beta_v$} \\
\noalign{\smallskip}
 0.10 & 0.010 &        10.00 &  0.65 &  9.79 &  0.184 &  0.40 & 0.033 & $\,$    0.05 &  0.24 &  3.61 &  0.193 &  0.50 & 0.150 & $\,$    0.33 &  0.33 &  1.22 & -0.106 \\
 0.10 & 0.033 &        10.00 &  0.35 &  8.17 & -0.143 &  0.40 & 0.033 & $\,$    0.10 &  0.28 &  3.46 &  0.214 &  0.50 & 0.150 & $\,$    1.00 &  0.55 &  1.37 & -0.138 \\
 0.15 & 0.010 & $\;$    3.00 &  0.49 &  8.26 &  0.305 &  0.40 & 0.033 & $\,$    0.33 &  0.50 &  2.83 &  0.207 &  0.50 & 0.150 & $\,$    3.00 &  0.60 &  1.41 & -0.117 \\
 0.15 & 0.010 &        10.00 &  0.87 &  6.38 &  0.053 &  0.40 & 0.033 & $\,$    1.00 &  0.79 &  2.30 &  0.078 &  0.50 & 0.150 &        10.00 &  0.61 &  1.44 & -0.115 \\
 0.15 & 0.033 & $\;$    3.00 &  0.27 &  6.07 & -0.124 &  0.40 & 0.033 & $\,$    3.00 &  0.93 &  2.39 &  0.004 &  0.50 & 0.200 & $\,$    1.00 &  0.24 &  1.05 & -0.152 \\
 0.15 & 0.033 &        10.00 &  0.77 &  5.93 & -0.046 &  0.40 & 0.033 &        10.00 &  0.97 &  2.45 &  0.027 &  0.50 & 0.200 & $\,$    3.00 &  0.24 &  1.07 & -0.100 \\
 0.15 & 0.060 &        10.00 &  0.50 &  5.09 & -0.150 &  0.40 & 0.060 & $\,$    0.00 &  0.11 &  2.34 & -0.135 &  0.50 & 0.200 &        10.00 &  0.28 &  1.08 & -0.171 \\
 0.20 & 0.010 & $\;$    1.00 &  0.22 &  8.91 &  0.418 &  0.40 & 0.060 & $\,$    0.05 &  0.16 &  2.30 & -0.095 &  0.75 & 0.010 & $\,$    0.00 &  0.99 &  1.42 &  0.092 \\
 0.20 & 0.010 & $\;$    3.00 &  0.75 &  5.01 &  0.180 &  0.40 & 0.060 & $\,$    0.10 &  0.16 &  2.35 & -0.064 &  0.75 & 0.010 & $\,$    0.05 &  0.99 &  1.40 &  0.070 \\
 0.20 & 0.010 &        10.00 &  0.94 &  4.73 &  0.040 &  0.40 & 0.060 & $\,$    0.33 &  0.36 &  2.31 &  0.042 &  0.75 & 0.010 & $\,$    0.10 &  0.99 &  1.40 &  0.075 \\
 0.20 & 0.033 & $\;$    3.00 &  0.66 &  4.50 &  0.053 &  0.40 & 0.060 & $\,$    1.00 &  0.72 &  2.15 &  0.030 &  0.75 & 0.010 & $\,$    0.33 &  0.99 &  1.37 &  0.048 \\
 0.20 & 0.033 &        10.00 &  0.89 &  4.57 &  0.008 &  0.40 & 0.060 & $\,$    3.00 &  0.89 &  2.23 & -0.003 &  0.75 & 0.010 & $\,$    1.00 &  0.99 &  1.34 &  0.036 \\
 0.20 & 0.060 & $\;$    3.00 &  0.48 &  3.81 & -0.114 &  0.40 & 0.060 &        10.00 &  0.93 &  2.35 & -0.017 &  0.75 & 0.010 & $\,$    3.00 &  1.00 &  1.34 &  0.034 \\
 0.20 & 0.060 &        10.00 &  0.75 &  4.16 & -0.088 &  0.40 & 0.100 & $\,$    0.33 &  0.24 &  1.76 & -0.148 &  0.75 & 0.010 &        10.00 &  1.00 &  1.33 &  0.050 \\
 0.25 & 0.010 & $\;$    1.00 &  0.52 &  5.11 &  0.331 &  0.40 & 0.100 & $\,$    1.00 &  0.60 &  1.85 & -0.078 &  0.75 & 0.033 & $\,$    0.00 &  0.97 &  1.42 &  0.076 \\
 0.25 & 0.010 & $\;$    3.00 &  0.85 &  3.80 &  0.086 &  0.40 & 0.100 & $\,$    3.00 &  0.75 &  2.05 & -0.065 &  0.75 & 0.033 & $\,$    0.05 &  0.97 &  1.40 &  0.074 \\
 0.25 & 0.010 &        10.00 &  0.96 &  3.81 &  0.002 &  0.40 & 0.100 &        10.00 &  0.76 &  2.08 & -0.071 &  0.75 & 0.033 & $\,$    0.10 &  0.97 &  1.40 &  0.067 \\
 0.25 & 0.033 & $\;$    1.00 &  0.40 &  4.10 &  0.145 &  0.40 & 0.150 & $\,$    1.00 &  0.35 &  1.54 & -0.153 &  0.75 & 0.033 & $\,$    0.33 &  0.98 &  1.38 &  0.065 \\
 0.25 & 0.033 & $\;$    3.00 &  0.80 &  3.55 &  0.023 &  0.40 & 0.150 & $\,$    3.00 &  0.42 &  1.63 & -0.164 &  0.75 & 0.033 & $\,$    1.00 &  0.99 &  1.31 &  0.041 \\
 0.25 & 0.033 &        10.00 &  0.94 &  3.70 & -0.004 &  0.40 & 0.150 &        10.00 &  0.45 &  1.70 & -0.157 &  0.75 & 0.033 & $\,$    3.00 &  0.99 &  1.34 &  0.044 \\
 0.25 & 0.060 & $\;$    1.00 &  0.25 &  3.14 & -0.116 &  0.50 & 0.010 & $\,$    0.00 &  0.72 &  2.95 &  0.269 &  0.75 & 0.033 &        10.00 &  0.99 &  1.31 &  0.040 \\
 0.25 & 0.060 & $\;$    3.00 &  0.68 &  3.25 & -0.060 &  0.50 & 0.010 & $\,$    0.05 &  0.71 &  3.01 &  0.279 &  0.75 & 0.060 & $\,$    0.00 &  0.94 &  1.33 &  0.070 \\
 0.25 & 0.060 &        10.00 &  0.83 &  3.43 & -0.050 &  0.50 & 0.010 & $\,$    0.10 &  0.73 &  2.86 &  0.264 &  0.75 & 0.060 & $\,$    0.05 &  0.94 &  1.32 &  0.058 \\
 0.25 & 0.100 & $\;$    3.00 &  0.41 &  2.78 & -0.123 &  0.50 & 0.010 & $\,$    0.33 &  0.82 &  2.36 &  0.199 &  0.75 & 0.060 & $\,$    0.10 &  0.94 &  1.34 &  0.070 \\
 0.25 & 0.100 &        10.00 &  0.47 &  2.89 & -0.156 &  0.50 & 0.010 & $\,$    1.00 &  0.93 &  1.98 &  0.073 &  0.75 & 0.060 & $\,$    0.33 &  0.95 &  1.33 &  0.061 \\
 0.33 & 0.010 & $\;$    0.00 &  0.01 &  6.91 &  0.254 &  0.50 & 0.010 & $\,$    3.00 &  0.98 &  1.95 &  0.041 &  0.75 & 0.060 & $\,$    1.00 &  0.96 &  1.30 &  0.035 \\
 0.33 & 0.010 & $\;$    0.05 &  0.01 &  7.53 & -0.417 &  0.50 & 0.010 &        10.00 &  0.99 &  1.96 &  0.009 &  0.75 & 0.060 & $\,$    3.00 &  0.97 &  1.28 &  0.027 \\
 0.33 & 0.010 & $\;$    0.10 &  0.04 &  8.66 &  0.374 &  0.50 & 0.033 & $\,$    0.00 &  0.64 &  2.64 &  0.221 &  0.75 & 0.060 &        10.00 &  0.97 &  1.28 &  0.013 \\
 0.33 & 0.010 & $\;$    0.33 &  0.35 &  5.00 &  0.393 &  0.50 & 0.033 & $\,$    0.05 &  0.65 &  2.59 &  0.228 &  0.75 & 0.100 & $\,$    0.00 &  0.86 &  1.23 &  0.011 \\
 0.33 & 0.010 & $\;$    1.00 &  0.74 &  3.00 &  0.165 &  0.50 & 0.033 & $\,$    0.10 &  0.67 &  2.41 &  0.207 &  0.75 & 0.100 & $\,$    0.05 &  0.86 &  1.20 &  0.017 \\
 0.33 & 0.010 & $\;$    3.00 &  0.92 &  2.84 &  0.028 &  0.50 & 0.033 & $\,$    0.33 &  0.77 &  2.16 &  0.152 &  0.75 & 0.100 & $\,$    0.10 &  0.82 &  1.26 &  0.037 \\
 0.33 & 0.010 &        10.00 &  0.98 &  2.97 & -0.002 &  0.50 & 0.033 & $\,$    1.00 &  0.90 &  1.87 &  0.049 &  0.75 & 0.100 & $\,$    0.33 &  0.87 &  1.20 &  0.019 \\
 0.33 & 0.033 & $\;$    0.33 &  0.25 &  3.54 &  0.166 &  0.50 & 0.033 & $\,$    3.00 &  0.96 &  1.91 &  0.022 &  0.75 & 0.100 & $\,$    1.00 &  0.89 &  1.20 & -0.023 \\
 0.33 & 0.033 & $\;$    1.00 &  0.68 &  2.79 &  0.122 &  0.50 & 0.033 &        10.00 &  0.98 &  1.94 &  0.013 &  0.75 & 0.100 & $\,$    3.00 &  0.91 &  1.18 & -0.015 \\
 0.33 & 0.033 & $\;$    3.00 &  0.89 &  2.76 &  0.028 &  0.50 & 0.060 & $\,$    0.00 &  0.52 &  2.13 &  0.103 &  0.75 & 0.100 &        10.00 &  0.91 &  1.22 & -0.011 \\
 0.33 & 0.033 &        10.00 &  0.96 &  2.92 & -0.003 &  0.50 & 0.060 & $\,$    0.05 &  0.55 &  2.07 &  0.093 &  0.75 & 0.150 & $\,$    0.00 &  0.70 &  1.05 & -0.057 \\
 0.33 & 0.060 & $\;$    0.33 &  0.15 &  2.48 & -0.064 &  0.50 & 0.060 & $\,$    0.10 &  0.55 &  2.11 &  0.107 &  0.75 & 0.150 & $\,$    0.05 &  0.69 &  1.06 & -0.054 \\
 0.33 & 0.060 & $\;$    1.00 &  0.59 &  2.50 &  0.022 &  0.50 & 0.060 & $\,$    0.33 &  0.67 &  1.98 &  0.087 &  0.75 & 0.150 & $\,$    0.10 &  0.71 &  1.02 & -0.038 \\
 0.33 & 0.060 & $\;$    3.00 &  0.84 &  2.61 & -0.031 &  0.50 & 0.060 & $\,$    1.00 &  0.84 &  1.79 &  0.018 &  0.75 & 0.150 & $\,$    0.33 &  0.73 &  1.05 & -0.040 \\
 0.33 & 0.060 &        10.00 &  0.90 &  2.77 & -0.021 &  0.50 & 0.060 & $\,$    3.00 &  0.93 &  1.84 &  0.014 &  0.75 & 0.150 & $\,$    1.00 &  0.77 &  1.05 & -0.036 \\
 0.33 & 0.100 & $\;$    1.00 &  0.44 &  2.13 & -0.149 &  0.50 & 0.060 &        10.00 &  0.95 &  1.88 & -0.003 &  0.75 & 0.150 & $\,$    3.00 &  0.78 &  1.06 & -0.060 \\
 0.33 & 0.100 & $\;$    3.00 &  0.67 &  2.33 & -0.097 &  0.50 & 0.100 & $\,$    0.00 &  0.35 &  1.57 & -0.081 &  0.75 & 0.150 &        10.00 &  0.78 &  1.05 & -0.044 \\
 0.33 & 0.100 &        10.00 &  0.68 &  2.39 & -0.093 &  0.50 & 0.100 & $\,$    0.05 &  0.35 &  1.59 & -0.072 &  0.75 & 0.200 & $\,$    0.00 &  0.50 &  0.84 & -0.115 \\
 0.40 & 0.010 & $\;$    0.00 &  0.34 &  4.84 &  0.392 &  0.50 & 0.100 & $\,$    0.10 &  0.39 &  1.61 & -0.057 &  0.75 & 0.200 & $\,$    0.05 &  0.52 &  0.83 & -0.104 \\
 0.40 & 0.010 & $\;$    0.05 &  0.35 &  5.22 &  0.423 &  0.50 & 0.100 & $\,$    0.33 &  0.52 &  1.56 & -0.038 &  0.75 & 0.200 & $\,$    0.10 &  0.50 &  0.85 & -0.110 \\
 0.40 & 0.010 & $\;$    0.10 &  0.37 &  4.77 &  0.393 &  0.50 & 0.100 & $\,$    1.00 &  0.73 &  1.58 & -0.042 &  0.75 & 0.200 & $\,$    0.33 &  0.54 &  0.86 & -0.109 \\
 0.40 & 0.010 & $\;$    0.33 &  0.59 &  3.41 &  0.302 &  0.50 & 0.100 & $\,$    3.00 &  0.83 &  1.71 & -0.036 &  0.75 & 0.200 & $\,$    1.00 &  0.57 &  0.88 & -0.082 \\
 0.40 & 0.010 & $\;$    1.00 &  0.84 &  2.47 &  0.106 &  0.50 & 0.100 &        10.00 &  0.84 &  1.69 & -0.056 &  0.75 & 0.200 & $\,$    3.00 &  0.59 &  0.90 & -0.083 \\
 0.40 & 0.010 & $\;$    3.00 &  0.95 &  2.41 &  0.031 &  0.50 & 0.150 & $\,$    0.00 &  0.13 &  1.11 & -0.128 &  0.75 & 0.200 &        10.00 &  0.59 &  0.89 & -0.096 \\
 0.40 & 0.010 &        10.00 &  0.99 &  2.47 &  0.014 &  0.50 & 0.150 & $\,$    0.05 &  0.14 &  1.14 & -0.165 & \\
 0.40 & 0.033 & $\;$    0.00 &  0.25 &  3.46 &  0.220 &  0.50 & 0.150 & $\,$    0.10 &  0.14 &  1.13 & -0.179 & \\
\noalign{\smallskip}
\end{tabular}
\end{table*}

Table~1 gives an overview of the simulations performed. 
The first three columns give the star formation 
efficiency $\epsilon$, the ratio of $r_h/r_t$, and the value of $\tau_M/t_{Cross}$. The next columns give the fraction of
bound mass remaining at the end of the runs $fst=M_{* f}/M_{ecl}$, the final half-mass radius in terms of 
the initial one and the global anisotropy parameter $\beta_v$, which is defined by the following formula:
\begin{equation}
\beta_v = 1 - \frac{\sum_i v^2_t}{2 \sum_i v^2_r} \;\; .
\label{eq:aniso}
\end{equation}
Here the sums run over all stars bound to the star cluster at the end of the runs and $v_t$ and $v_r$ are
the tangential and radial velocity component of each star. The anisotropies were calculated after the stellar
velocities were transformed back into a non-rotating coordinate system. For isotropic velocity
dispersions, $\beta_v=0$, while radial (tangential) anisotropic
velocity dispersions correspond to positive (negative) values of $\beta_v$.
A complete list of results, including the time evolution of different parameters and data for runs
which do not lead to cluster survival can be found under the following internet address:
http://www.astro.uni-bonn.de/$\sim$webaiub/german/downloads.php/ .

\section{Results}
\label{sec:results}

\subsection{Final cluster properties in dependence of initial parameters}

\begin{figure*}
\begin{center}
\includegraphics[width=17cm]{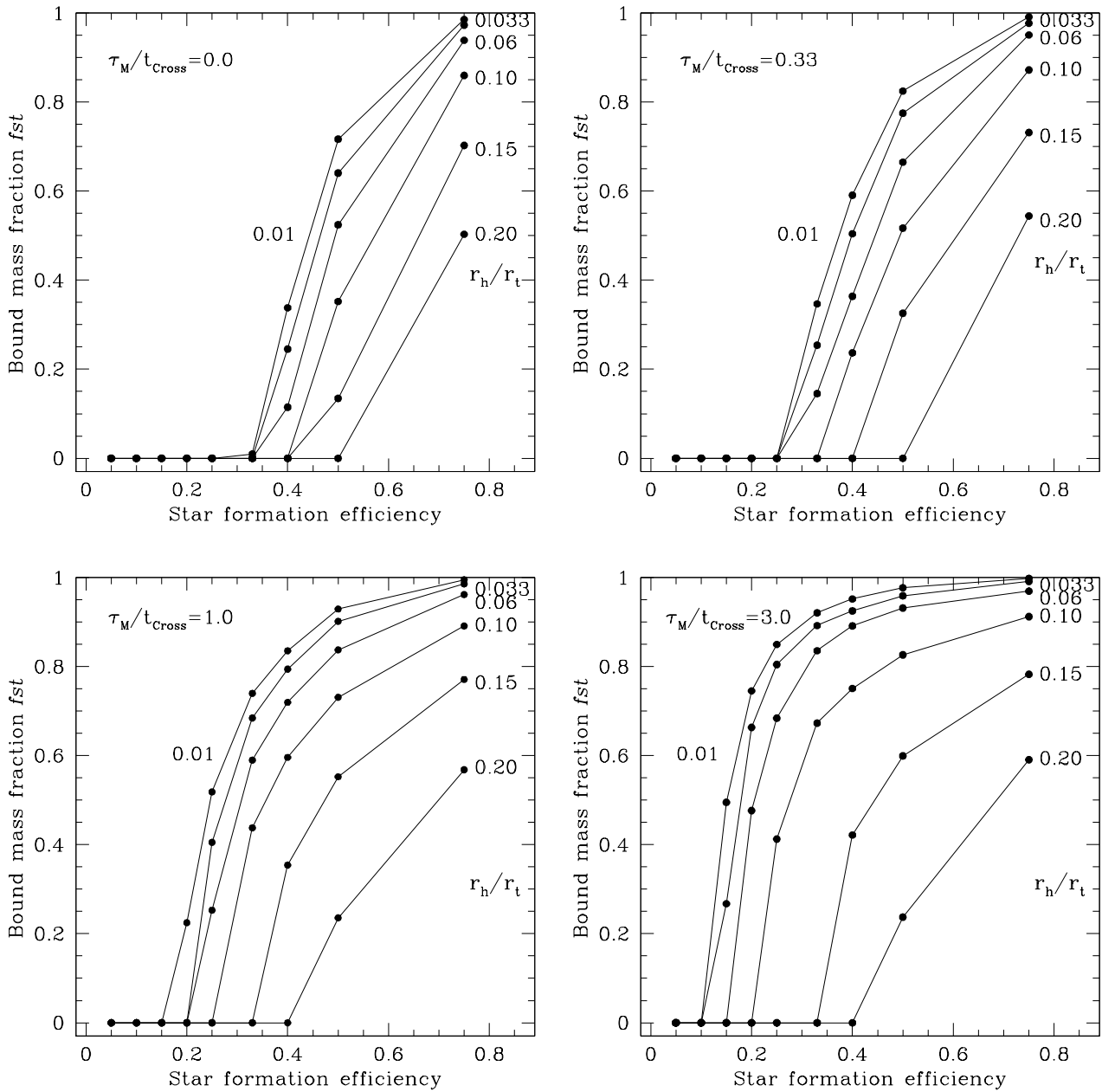}
\end{center}
\caption{Surviving mass fraction as a function of the star formation efficiency. Curves are labeled by
the ratio of $r_h/r_t$ used in the runs. The upper left panel shows runs in which the gas was removed
instantaneously ($\tau_M/t_{Cross}=0.0$). In such a case SFEs larger than 33\% are necessary to produce
bound clusters. The other panels show runs in which the gas was removed on longer timescales:
$\tau_M/t_{Cross}=0.33$ (upper right), $\tau_M/t_{Cross}=1.00$ (lower left) and $\tau_M/t_{Cross}=3.00$
(lower right). In these cases a larger number of stars remain bound to the clusters. In all
cases, clusters in a stronger tidal field are more easily destroyed.}
\label{fig:bound:sfe}
\end{figure*}

\begin{figure*}
\begin{center}
\includegraphics[width=17cm]{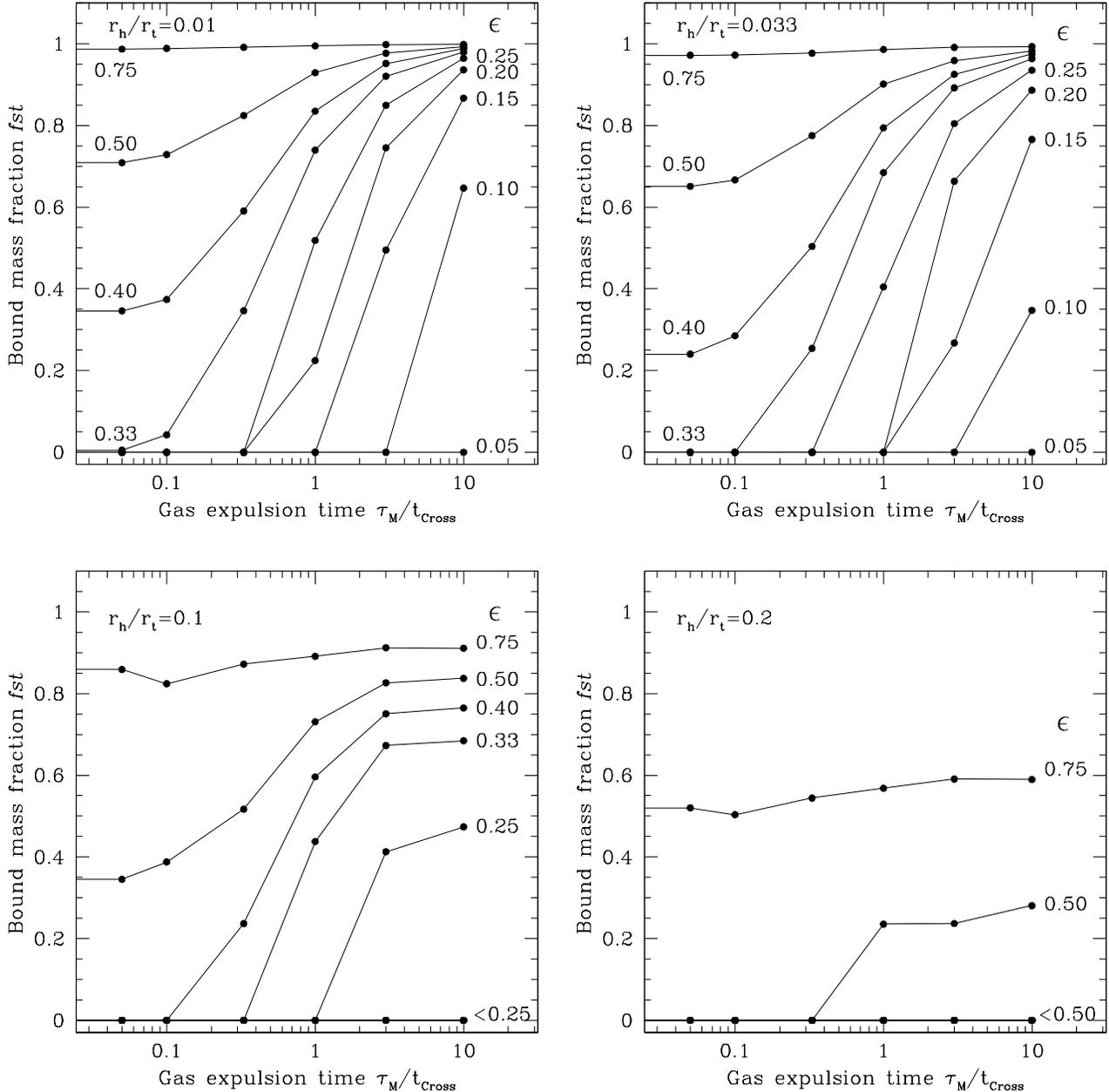}
\end{center}
\caption{Same as Fig. \ref{fig:bound:sfe} but now in dependence of the gas expulsion timescale. Curves are labeled
by the assumed star formation efficiency. Gas expulsion timescales below $\tau_M/t_{Cross}=0.1$ give nearly
identical results. In clusters formed in strong tidal fields ($r_h/r_t=0.20$), star formation efficiencies 
$\gtrsim 50$\% are needed in order to produce a bound system.}
\label{fig:bound:gas}
\end{figure*}

Figs.\ \ref{fig:bound:sfe} and \ref{fig:bound:gas} depict results for the bound mass fraction at the end of the
simulations as a function of star
formation efficiency, strength of the external tidal field and gas expulsion timescale. The different
panels in Fig.\ \ref{fig:bound:sfe} depict four cases for each of which the ratio of gas expulsion timescale to the 
crossing time of the clusters was held constant.
In case of instantaneous gas-loss (upper left panel), models with star formation efficiencies of 40\%
or larger result in surviving clusters. In case of a very weak external tidal field ($r_h/r_t=0.01$), even the
model with a SFE of 33\% produces a bound cluster containing 0.9\% of the initial stars, which agrees
will with estimates from the literature \citep{lmd84}.

Prolonging the timescale on which the gas is removed also increases the mass fraction of bound stars in the 
clusters.
If the tidal field is weak, star clusters can survive star
formation efficiencies below 20\% if the gas is removed slowly (see Table~1 and uppermost 
curves in the lower right panel). \citet{kah01} performed fully realistic $N$-body calculations which included a 
tidal field realistic for the solar neighbourhood and gas expulsion on a thermal time scale. For
their model B, which had $\tau_M/t_{Cross}=0.3$ and $r_h/r_t=0.01$ and a SFE of 33\%, they found a
bound mass fraction of 25\% after gas expulsion, which agrees very well with the results of our runs.
For slow gas removal, nearly 100\% of stars remain bound if the star formation efficiency
is larger than 30\%. 

Our simulations also show that the limit for survival depends on the
strength of the external tidal field: The higher the ratio of $r_h/r_t$, the higher is the star formation
efficiency needed to produce a surviving cluster. As long as the strength of the  external tidal field
is smaller than $r_h/r_t \le 0.06$, it has only a moderate influence on cluster dissolution. However, cluster 
survivability is drastically reduced if the strength of the external tidal field is changed from 
$r_h/r_t=0.1$ to $r_h/r_t=0.2$. In the case of $r_h/r_t=0.2$, SFEs of less than 50\% 
do not lead to the formation a bound star cluster since expanding clusters get easily disrupted due to the 
strong external tidal field; in this case a small initial mass loss leads to an expansion
which triggers catastrophic mass-loss across the tidal boundary. Star clusters must therefore form 
concentrated in order to survive gas expulsion. This is possible if they form for example as either
high concentration ($c \gtrsim 1.5$) King models or underfilling their tidal radius.

Similar results can also be found in Fig.\ \ref{fig:bound:gas}, which depicts the bound mass fraction as a
function of the gas expulsion timescale. For gas expulsion timescales up to and including $\tau_M/t_{Cross}=0.10$, 
our results are nearly independent of the actual value of $\tau_M/t_{Cross}$, showing that as long as the 
gas is removed with $\tau_M \ll t_{Cross}$, the results do not depend on the details of the gas removal. 
Similarly, differences between the $\tau_M/t_{Cross}=3.0$ and the $\tau_M/T_{Cross}=10.0$ case are also relatively small,
at least for high SFEs.

For a typical star formation efficiency of 25\%, the gas expulsion timescale has to be larger than
$\tau_M/t_{Cross}=0.33$ in order to produce a bound star cluster. In our runs, star formation efficiencies 
of 5\% did not lead to the formation of bound star clusters, which justifies that we did not perform any runs with lower SFE.

\begin{figure}
\begin{center}
\includegraphics[width=8.3cm]{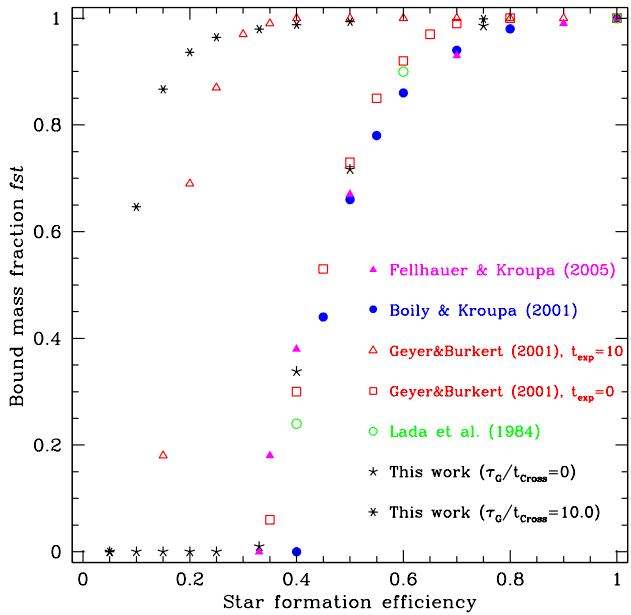}
\end{center}
\caption{Comparison of the surviving mass fraction derived in this
work with results from the literature. For instantaneous gas removal (right line of points) there is
very good agreement between both. For slow gas removal 
(left points, stars and open triangles), the
critical SFE needed to produce a bound cluster determined here is about 5\% smaller than the one found by \citet{gb01}.
This can be explained by the different initial density profiles and the fact that \citet{gb01} assumed linear gas removal 
while we assume an exponential one.}
\label{fig:comp}
\end{figure}

Figs.\ \ref{fig:comp} compares our results for the bound mass fraction as a function of the star
formation efficiency with published results from the literature. Shown are cases when the gas is removed
instantaneously (right group of points) and cases of slow gas removal. It can be seen that
for instantaneous gas removal there is very good agreement between the results of this paper and published
results. SFEs of 33\% already lead to a final bound cluster, although only a very small mass fraction
remains bound in this case. For a SFE of 50\%, about 70\% of the total cluster mass remains bound.
In case of near adiabatic gas removal, we find that the critical SFE needed to produce a bound cluster is
between 5\% to 10\%. This is about 5\% smaller than what \citet{gb01} found for their model N2 with $t_{exp}=10$. 
Performing additional $N$-body runs shows that the difference becomes significantly smaller if we let the gas 
fraction decrease linearly with time, as was done by \citet{gb01}. The remaining difference is probably
due to the different density profiles. \citet{gb01} used King $W_0=3$ and $W_0=5$ models in their runs, which
are significantly less concentrated than the Plummer models we use.

\begin{figure*}
\begin{center}
\includegraphics[width=17cm]{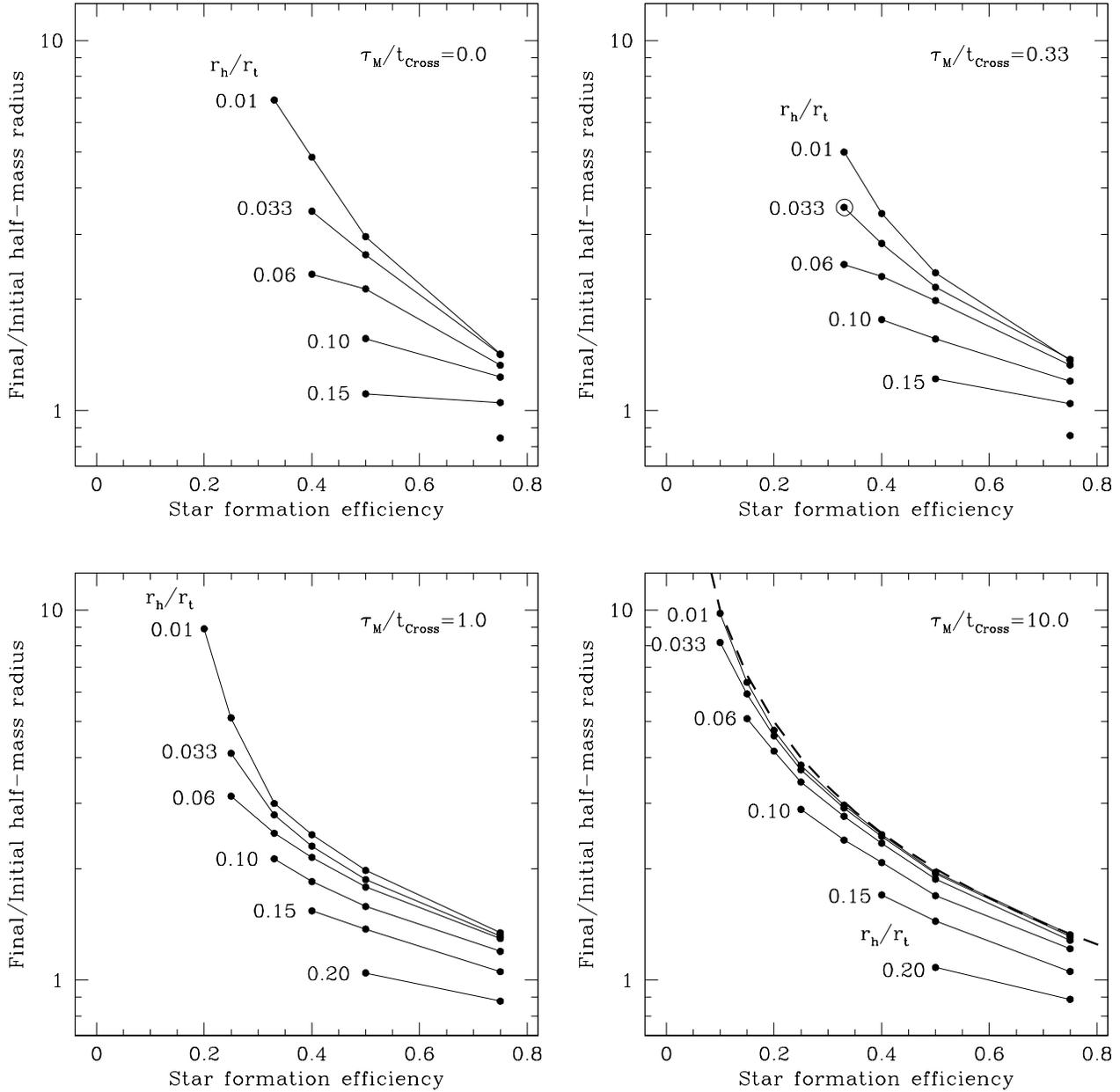}
\end{center}
\caption{Same as Fig. \ref{fig:bound:sfe} for the ratio of the final half-mass radius to the initial one.
The curves are labeled by the ratio of $r_h/r_t$ used.
Clusters with lower star formation efficiencies generally expand stronger than those with higher ones.
Largest expansion factors are of order 10 and are reached for clusters which are nearly dissolved. Typical
expansion factors are around 3 or 4 for SFEs of 25\%, which means that open/globular clusters should have formed
more concentrated by such a factor than as we see them now. Expansion factors are smaller for clusters in stronger
tidal fields due to the efficient removal of stars at larger radii, and for adiabatic mass-loss.
The open circle in the upper right panel indicates the cluster discussed in Figs. \ref{fig:radall}
to \ref{fig:anisoi}. The dashed line in the lower right panel shows the theoretical prediction for the expansion
factor in case of adiabatic gas loss (eq.\ \ref{eq:exp}).}
\label{fig:finrad}
\end{figure*}

Fig.\ \ref{fig:finrad} depicts the ratio of the final half-mass radius compared to the initial one as a function
of the star formation efficiency. In contrast to the previous two figures, only models leading to bound clusters 
are plotted, which explains why the curves don't extend to low star formation efficiencies. 

Due to the lowering of the cluster potential when the gas is removed, all star clusters
expand and the final half-mass radii are mostly larger than the initial ones. This is despite the fact that
the tidal field removes weakly bound stars from the clusters. If gas is removed adiabatically,
the total energy is conserved and the ratio of final to initial half-mass radius must obey the following relation
\citep{hill80}:
\begin{equation}
\frac{r_{h f}}{r_{h i}} = \frac{M_{ecl}}{M_{* f}}
\label{eq:exp}
\end{equation}
The dashed line in the lower right panel shows the expected expansion due to eq.\ \ref{eq:exp}. It indeed gives a good fit to
our results for nearly isolated clusters in which the gas is removed slowly.

Clusters in 
strong tidal fields ($r_h/r_t \ge 0.15$) show only very small expansion as a result of the gas
removal. For typical values of the star formation efficiency ($\approx$25\%), expansion 
factors are around 3 or 4 in most cases. Hence, typical half-mass radii of embedded star clusters 
should be smaller by
this factor than the corresponding ones of open or globular clusters. Observational data indeed suggests expansion,
although the expansion factors may be even larger than found here \citep{k05}. The reason could be that
shortly after gas expulsion, unbound stars are still sufficiently close to the cluster to be counted
as members, making the clusters appear larger. In addition, young cluster systems contain a mix of bound clusters 
and more extended clusters which are dissolving but are still sufficiently compact to appear as star 
clusters. Both effects might increase the observed average cluster radius for systems which are a
few 10s of Myr old (see discussion of Fig.\ \ref{fig:radall} and \citet{bg06}). 

Fig.\ \ref{fig:finrad} also shows that in each model family, the expansion is strongest for clusters which
have the smallest star formation efficiency that still produces a bound star cluster. This is to be expected
since low star formation efficiency means a stronger decrease of the overall cluster potential and 
as long as the tidal field is weak, a cluster can expand since stars in the halo are not removed. For specific parameter
combinations, the final half-mass radius can be a factor 10 higher than the initial one. It can also be
seen from Fig.\ \ref{fig:finrad} that quick mass-loss generally leads to stronger expansion than adiabatic 
mass-loss.

\begin{figure}
\begin{center}
\includegraphics[width=8.3cm]{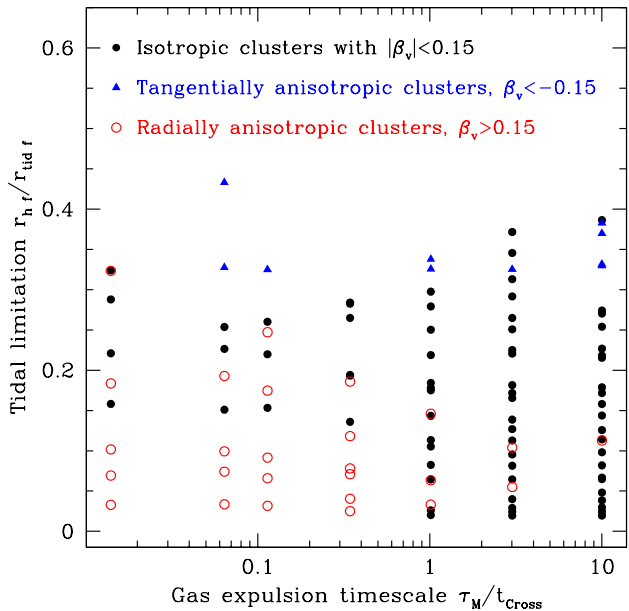}
\end{center}
\caption{Global anisotropy $\beta_v$ for the final clusters with $\epsilon\le0.50$ as a function of the gas expulsion timescale and the
ratio of the final half-mass radius to the tidal radius. Clusters in which the gas was removed quickly
have radially anisotropic velocity distributions due to the sudden decrease in cluster
potential (open circles). Adiabatic gas-loss leads to more or less isotropic velocity dispersions 
(filled circles), at least
if the clusters also start isotropic.
The exception are a few clusters in strong tidal fields which have acquired tangentially isotropic velocity dispersions
(triangles) due to the tidal field. These clusters are also rotating due to the external tidal field.}
\label{fig:aniso}
\end{figure}

Fig.\ \ref{fig:aniso} finally shows the global anisotropies of the clusters by the time the simulations were
stopped. Only clusters with $\epsilon \le 0.50$ are shown since models with higher SFEs are hardly affected 
by gas expulsion. The anisotropies were calculated using eq.\ \ref{eq:aniso} and summing over all stars
which were still bound to the clusters at the end of the simulations. Clusters in which the gas is 
removed quickly generally acquire radially anisotropic velocity 
distributions. This is due to the fact that such clusters are super-virial after gas removal. As a result,
the stars expand more or less radially outwards and the cluster halos become populated by stars on radial
orbits. A close inspection of these clusters shows that the innermost parts normally stay isotropic. 
Adiabatic gas-loss on the other hand leads to more or less isotropic velocity dispersions (filled circles
in Fig.\ \ref{fig:aniso}), at least if the clusters start with an isotropic velocity dispersion as is the case
in our runs.
If clusters are immersed in strong tidal fields (upper points in Fig.\ \ref{fig:aniso}), the global
velocity profile becomes tangentially anisotropic since the Coriolis force due to the tidal field forces 
the expanding cluster stars onto more tangential orbits. The clusters also acquire significant rotation
in their outer parts in this case as a result of the non-radial acceleration of the stars due to the Coriolis force, 
and possibly also angular momentum dependent escape of stars.

The features just described might be detectable in star clusters which are dynamically young, i.e. for which
the ratio of their relaxation time to their age is large $t_{Rel}/t_{Age} \ge 1$. A number of globular clusters
in the Milky Way, like for example $\omega$ Cen, have relaxation times of the order of a Hubble time or only
slightly smaller. For a subset of them, the tidal radius is also significantly larger than the cluster's half-mass
radius, meaning that tidal effects are likely not important for these clusters. Observation of the velocity
dispersion profile would allow to constrain the timescale over which the gas was expelled in these clusters.

\subsection{Evolution of individual clusters and observational consequences of gas expulsion}

\begin{figure*}
\begin{center}
\includegraphics[width=17.0cm]{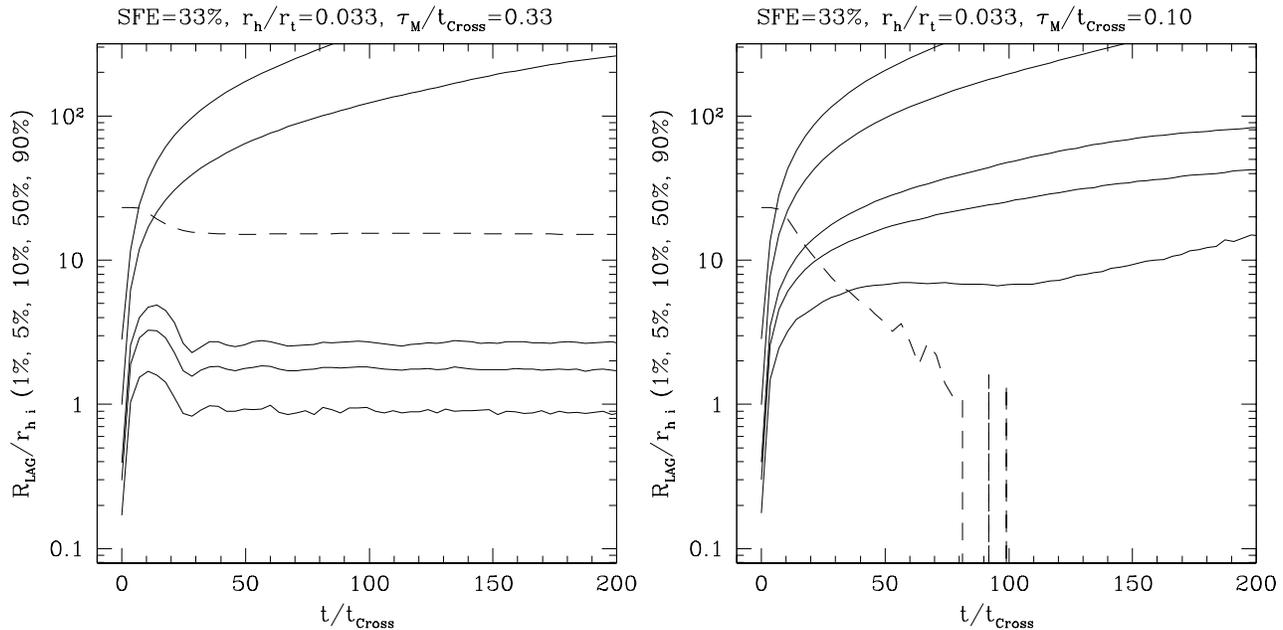}
\end{center}
\caption{Evolution of Lagrangian radii of all stars, including unbound ones, in terms of the initial half-mass radius for 
two star clusters as
a function of time, expressed in terms of the initial crossing time. The parameters of the clusters are
given at the top of the panels. Solid lines show evolution of Lagrangian radii for all stars including
unbound ones,
dashed lines the evolution of the tidal radius of all stars still bound to the clusters. The
cluster in the left panel expands strongly during the first 20 crossing times
in response to gas loss, but a fraction of its stars falls back later and
forms a bound star cluster. Gas expulsion happens slightly faster for the cluster in
the right panel, which is enough to unbind it completely.}
\label{fig:radall}
\end{figure*}

Fig.\ \ref{fig:radall} depicts the evolution of Lagrangian radii of all stars including unbound ones with time 
for two typical star clusters
from our runs. As a result of rapid gas expulsion, both star clusters expand strongly in the beginning,
pushing a large fraction of their stars over the tidal radius. For the cluster in the left panel, a
small fraction of stars falls back after about 10 to 20 initial crossing times and forms a bound cluster.
In this cluster, the Lagrangian radii become essentially constant with time
after about 50 initial crossing times, justifying our maximum simulation time of 350 initial crossing times. Compared
to the initial cluster, the
half-mass radius of the final cluster has increased by a factor 2.8 and only 32\% of the stars of the initial
cluster are still bound to the final cluster.

\begin{figure}
\begin{center}
\includegraphics[width=8.3cm]{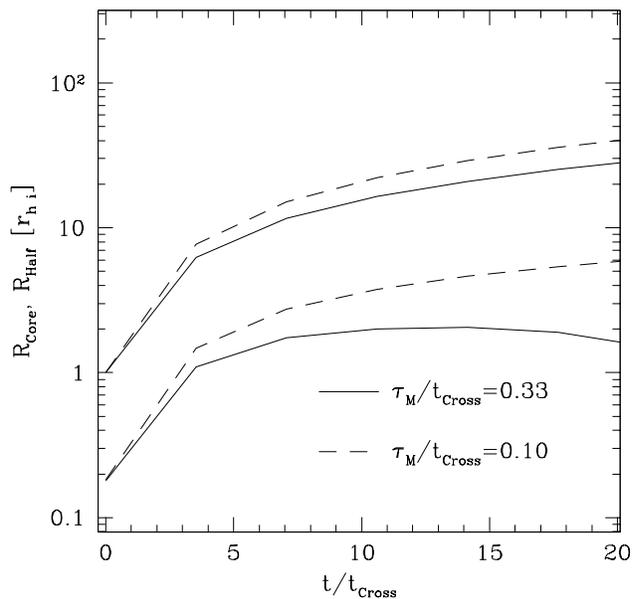}
\end{center}
\caption{Evolution of projected core (lower two curves) and half-mass radius (upper two curves) of all stars
of the two clusters depicted in Fig.\ \ref{fig:radall}. Although only the run with $\tau_M/t_{Cross}=0.33$
results in a bound cluster, core and half-mass radius evolve in a very similar way in both clusters
during the first 10 to 20 crossing times, corresponding to about 10 Myr for typical cluster parameters
(i.e. initial velocity dispersion of 5 pc/Myr). For young star clusters, it can therefore be difficult to 
distinguish bound from unbound clusters on the basis of their size or concentration alone.}
\label{fig:struct}
\end{figure}

The cluster in the right panel has the same star formation efficiency and $r_h/r_t$ ratio as the cluster in the
left panel, but loses the gas on a slightly faster timescale ($\tau_M/t_{Cross}=0.10$ compared
to $\tau_M/t_{Cross}=0.33$). This small difference is enough to completely unbind this
cluster. The dissolution happens within the first 50 initial crossing times. Nevertheless, the appearance of
both clusters up to this point is very similar in terms of the structural parameters seen by an observer.
This is depicted in more detail in Fig.\ \ref{fig:struct} which shows the evolution of the core radius (taken
to be the 5\% projected Lagrangian radius) and the projected half-mass radius of the two clusters of
Fig.\ \ref{fig:radall}. Here again all stars including unbound ones were used to calculate the radii. It 
can be seen that the half-mass radii evolve in a very similar way initially, since
even after
20 initial crossing times the half-mass radius of the dissolving cluster is only 40\% larger than
that of the surviving cluster. The difference in the core radii is larger, but is still within a factor of 4
after 20 initial crossing times. Since for typical cluster parameters, one crossing time is of the order of a 
few 0.1 Myrs, it will be difficult to discriminate bound from unbound clusters in young star cluster systems.
Note also that the expansion factor of the projected radii is about 10 in both cases, being consistent with 
empirical data \citep{k05}.

\begin{figure}
\begin{center}
\includegraphics[width=8.3cm]{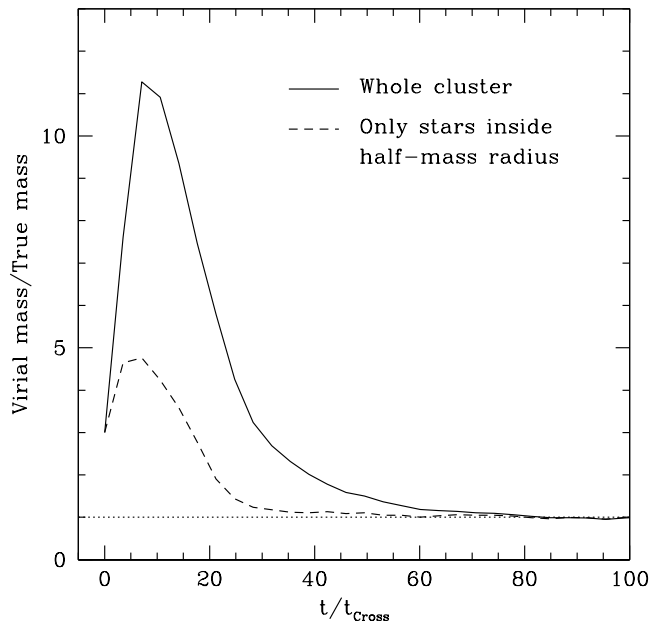}
\end{center}
\caption{Mass derived from the observed velocity dispersion compared to the true mass remaining
bound as a function of time for the surviving cluster of Figs.\ \ref{fig:radall} and \ref{fig:struct}.
Shown are mass estimates based on all stars inside the tidal radius (solid line) and inside the cluster's half-mass
radius (dashed line). Due to the assumed SFE of 33\%, the
initial total mass (stars plus gas) is three times higher than the stellar mass alone. The estimated mass
increases compared to the total mass due to gas removal and the loss of a large fraction of the cluster mass.
Depending on which stars are used to derive the total cluster mass, it takes between 30 to 50 initial crossing 
times (i.e. about 20 Myrs) until the virial mass becomes a reliable estimate for the total mass.}
\label{fig:virial}
\end{figure}

Fig.\ \ref{fig:virial} depicts the total mass that an observer would derive based on the velocity dispersion,
compared to the total mass remaining bound to the cluster. Virial masses were calculated for all stars projected
inside the tidal radius of the cluster and for all stars projected inside the half-mass radius.
The virial ratio is 3 initially since the cluster
formed with a SFE of 33\%. As the gas
is removed, many stars are pushed over the tidal radius and the bound mass drops. This explains the sharp rise
in the ratio of virial mass to the total bound mass during the first 10 crossing times. The virial ratio drops
later as the unbound stars leave the vicinity of the cluster. The initial increase is weaker if only stars inside
the half-mass radius are used to determine the virial mass due to the higher central density of the cluster and 
the smaller relative contribution
of escaping stars. Nevertheless, even inside the half-mass radius, it takes about 30 initial crossing times until
the virial mass estimate is a reliable estimate of the true cluster mass. This overestimate of a cluster's mass
due to the early violent relaxation associated with gas expulsion has recently also been pointed out by
\citet{bg06}.

\begin{figure}
\begin{center}
\includegraphics[width=8.3cm]{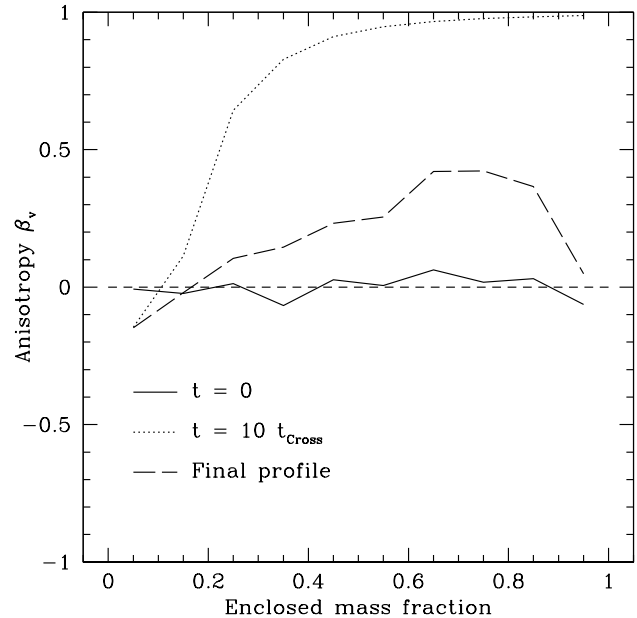}
\end{center}
\caption{Anisotropy parameter $\beta_v$ as a function of radius (expressed by the enclosed mass fraction) at three
different times for the same cluster as in Fig.\ \ref{fig:virial}. The initial cluster model (solid line) is isotropic
throughout.
In the beginning, the cluster becomes strongly radial in the outer parts due to gas removal which puts
many stars on outward-bound, ballistic orbits (dotted line). The final profile at the end of the run is nearly 
isotropic in the center and close to the tidal radius and mildly radially anisotropic at intermediate radii 
(dashed lines).}
\label{fig:anisoi}
\end{figure}

Fig.\ \ref{fig:anisoi} finally shows the anisotropy profile of the star cluster with $\epsilon$=0.33, $r_h/r_t=0.033$
and $\tau_M/t_{Cross}=0.33$. The anisotropy was calculated according to eq.\ \ref{eq:aniso}.
Our initial clusters are isotropic throughout. Due to the gas removal and the resulting escape of stars, the
velocity dispersion becomes strongly radially anisotropic in the outer parts. After 10 initial crossing times, the
anisotropy reaches nearly $\beta_v=1.0$ close to the tidal radius due to escaping stars. After the unbound stars 
have left the cluster, the radial
anisotropy decreases and for the cluster shown the final profile is isotropic again in the center and close to 
the tidal radius, and mildly radial anisotropic at intermediate radii. 

\section{Conclusions}
\label{sec:concl}

We have performed a large grid of simulations studying the impact of
initial gas expulsion on the survival rate and final properties of
star clusters, varying the star formation efficiency, ratio of gas
expulsion timescale to the crossing time of the cluster and the 
strength of the external tidal field.

Our simulations show that both the star formation efficiency
and the speed with which the gas is removed have a strong influence 
on the evolution of star clusters. In the case of instantaneous gas removal, 
clusters have to form with SFEs $\ge 33\%$ in order to survive gas expulsion.
This limit is significantly lowered for gas removal on longer timescales and clusters with 
SFEs as low as 10\% can survive gas expulsion in the adiabatic limit if the external tidal 
field is weak.
External tidal fields have a significant
influence on the cluster evolution only if the ratio of $r_h/r_t$ is
larger than about 0.05. Below this value, star clusters behave nearly as if
they are isolated. 

All star clusters expand due to gas expulsion. For star formation efficiencies around 25\%, star 
clusters expand by a factor of 3 or 4, nearly
independent of the speed with which the gas is removed. Open or globular clusters
must therefore have formed more compact and with higher central
densities than with what we see them today. Star clusters that just managed to
hold together should have expanded strongest and in rare
cases expansion factors of 10 or more are possible.

The velocity dispersion profiles of surviving clusters are tangentially 
anisotropic in their outer parts if the clusters are strongly effected by the external tidal field.
For nearly isolated clusters, the velocity profile is radially 
anisotropic if the gas was removed on a timescale smaller than the crossing
time of the cluster, while slow gas expulsion leads to mainly isotropic velocity
dispersions. In the central parts of star clusters, the velocity profile preserves 
its initial shape. This offers a chance to test how star formation and gas expulsion is 
happening in nature by studying the velocity profiles
of sufficiently isolated and dynamically young ($t_{Rel} > t_{Age}$) star clusters.

The simulations reported here should be useful for a number of follow-up projects.
First, they allow us to study the impact of gas expulsion on the evolution of the mass
function and other properties of whole globular cluster systems in galaxies: Starting with 
a range of cluster parameters and SFEs, the mass fraction remaining bound to each 
individual cluster can be calculated by interpolating between the runs in our grid. Repeating
the process for all clusters will then give the impact of gas expulsion on the whole
star cluster system and as a function of the age of the system. Our runs also
allow to test to which degree mass determinations of young star clusters through 
measurements of the radial velocity dispersion are affected by the initial gas 
expulsion \citep{bg06}. Finally, observations of individual star clusters
can be directly compared with our simulations to infer their starting conditions.
We plan to carry out such projects in the near future.

\section*{Acknowledgements}
We are grateful to Genevieve Parmentier and Ian Bonnell for useful discussions. This work was supported
by the DFG Priority Program 1177 'Witnesses of Cosmic History: Formation and evolution
of black holes, galaxies and their environment'.

\label{lastpage}


\begin{thebibliography}{}

\bibitem[\protect\citeauthoryear{Aarseth}{1985}]{a85}
Aarseth, S.~J., 1985, in {\it Multiple Time Scales}, Brackbill J.~U., 
  Cohen B.~I., eds., Academic Press, New York, p.\ 377

\bibitem[\protect\citeauthoryear{Aarseth}{1999}]{a99}
Aarseth, S.~J., 1999, PASP, 111, 1333

\bibitem[\protect\citeauthoryear{Adams}{2000}]{a00}
Adams, F., 2000, ApJ, 542, 964 

\bibitem[\protect\citeauthoryear{Bastian \& Goodwin}{2006}]{bg06}
Bastian, N., Goodwin, S. P., 2006, MNRAS, 369, 9

\bibitem[\protect\citeauthoryear{Binney \& Tremaine}{1987}]{bt87}
Binney, J., Tremaine, S., 1987, Galactic Dynamics, Princeton Univ. Press,
  Princeton 

\bibitem[\protect\citeauthoryear{Boily \& Kroupa}{2003a}]{bk03a}
Boily, C. M., Kroupa, P., 2003a, MNRAS, 338, 665

\bibitem[\protect\citeauthoryear{Boily \& Kroupa}{2003b}]{bk03b}
Boily, C. M., Kroupa, P., 2003b, MNRAS, 338, 673

\bibitem[\protect\citeauthoryear{Churchwell}{1997}]{c97}
Churchwell, E., 1997, ApJ, 479, 59 

\bibitem[\protect\citeauthoryear{Clark \& Bonnell}{2004}]{cb04}
Clark, P. C., Bonnell, I. A., 2004, MNRAS, 347, L36

\bibitem[\protect\citeauthoryear{Fellhauer \& Kroupa}{2005}]{fk05}
Fellhauer, M., Kroupa, P., 2005, ApJ, 630, 879

\bibitem[\protect\citeauthoryear{Geyer \& Burkert}{2001}]{gb01}
Geyer, M. P., Burkert, A., 2001, MNRAS, 323, 988

\bibitem[\protect\citeauthoryear{Goodwin}{1997a}]{good97a}
Goodwin, S. P., 1997a, MNRAS, 284, 785

\bibitem[\protect\citeauthoryear{Goodwin}{1997b}]{good97b}
Goodwin, S. P., 1997b, MNRAS, 286, 669

\bibitem[\protect\citeauthoryear{Heggie \& Mathieu}{1986}]{hm85}
Heggie, D. C., Mathieu, R. D., 1986, in {\it LNP Vol.\ 267: 
The Use of Supercomputers in Stellar Dynamics}, eds. P.~Hut and S.~McMillan, p.\ 233  

\bibitem[\protect\citeauthoryear{Hills}{1980}]{hill80}
Hills, J. G., 1980, ApJ, 235, 986

\bibitem[\protect\citeauthoryear{Hut \& Heggie}{2003}]{hh03}
Hut, P., Heggie, D., 2003, The gravitational million-body problem, Cambridge University Press, p. 7

\bibitem[\protect\citeauthoryear{Kroupa}{1995}]{k95}
Kroupa, P., 1995, MNRAS, 277, 1491

\bibitem[\protect\citeauthoryear{Kroupa}{2002}]{k02}
Kroupa, P., 2002, MNRAS, 330, 707

\bibitem[\protect\citeauthoryear{Kroupa}{2005}]{k05}
Kroupa, P., 2005, {\it The Fundamental Building Blocks of Galaxies}, in 
 Proceedings of the Gaia Symposium "The Three-Dimensional Universe with Gaia", Turon, C., O'Flaherty, K.S.,
  Perryman, M.A.C., eds., p.\ 629, astro-ph/0412069

\bibitem[\protect\citeauthoryear{Kroupa, Aarseth \& Hurley}{2001}]{kah01}
Kroupa, P., Aarseth, S., Hurley, J., 2001, MNRAS, 321, 707

\bibitem[\protect\citeauthoryear{Kroupa, Petr \& McCaughrean}{1999}]{kpm99}
Kroupa, P., Petr, M. G., McCaughrean, M. J., 1999, New Astronomy, 4, 495

\bibitem[\protect\citeauthoryear{Lada}{1999}]{l99}
Lada, E. A., 1999, {\it The Role of Embedded Clusters in Star Formation}, in 
 NATO ASIC Proc. 540: The Origin of Stars and Planetary Systems, Lada, C.J. \& Kylafis, N.D. eds., p.\ 441

\bibitem[\protect\citeauthoryear{Lada, Margulis \& Dearborn}{1984}]{lmd84}
Lada, C. J., Margulis, M., Dearborn, D., 1984, ApJ, 285, 141 

\bibitem[\protect\citeauthoryear{Lada \& Lada}{2003}]{ll03}
Lada, C. J., Lada, E. A., 2003, ARA\&A, 41, 57 

\bibitem[\protect\citeauthoryear{Makino et al.}{2003}]{mfkn03}
Makino, J., Fukushige, T., Koga, M., \& Namura, K., 2003, PASJ, 55, 1163

\bibitem[\protect\citeauthoryear{Palla et al.}{2005}]{petal05}	
Palla, F., Randich, S., Flaccomio, E., Pallavicini, R., 2005, ApJ, 626, 49

\bibitem[\protect\citeauthoryear{Tutukov}{1978}]{tut78}
Tutukov, A. V., 1978, ApJ, 70, 57 

\bibitem[\protect\citeauthoryear{Weidner \& Kroupa}{2005}]{wk05}
Weidner, C., Kroupa, P., 2005, ApJ, 625, 754

\bibitem[\protect\citeauthoryear{Whitmore et al.}{1999}]{wetal99}
Whitmore, B. C., et al., 1999, AJ, 118, 1551

\end{thebibliography}
\end{document}